\documentclass[letterpaper, 10 pt, conference]{ieeeconf}  

\usepackage{amssymb}
\usepackage{amsmath}
\usepackage{graphicx}
\usepackage{bm}
\IEEEoverridecommandlockouts                              
\title{\LARGE \bf
Guided Machine Learning for power grid segmentation 
}

\author{A. Marot, S. Tazi, B. Donnot, P. Panciatici (RTE R\&D)$^{1}$ 
\thanks{$^{1}$R\'eseau de Transport d'\'Electricit\'e (French TSO)}%
}

\usepackage{lastpage}

\usepackage{fancyhdr}

\begin{document}

\maketitle
\thispagestyle{empty}
\pagestyle{empty}

\begin{abstract}
The segmentation of large scale power grids into zones is crucial for control room operators when managing the grid complexity near real time. In this paper we propose a new method in two steps which is able to automatically do this segmentation, while taking into account the real time context, in order to help them handle shifting dynamics. Our method relies on a "guided" machine learning approach.
As a first step, we define and compute a task specific "Influence Graph" in a guided manner. We indeed simulate on a grid state chosen interventions, representative of our task of interest (managing active power flows in our case). 
For visualization and interpretation, we then build a higher representation of the grid relevant to this task by applying the graph community detection algorithm \textit{Infomap} on this Influence Graph.
To illustrate our method and demonstrate its practical interest, we apply it on commonly used systems, the IEEE-14 and IEEE-118.  We show promising and original interpretable results,  especially on the previously well studied RTS-96 system for grid segmentation. We eventually share initial investigation and results on a large-scale system, the French power grid, whose segmentation had a surprising resemblance with RTE's historical partitioning.
\end{abstract}




\section{INTRODUCTION}

Well-established power systems such as the French power grid are starting to experience a 
transition with a steep rise in complexity. This is due in part to the changing nature of the grid, with an end to the ever increasing total consumption. This shifts the way we traditionally develop the grid. While we used to expand it by building new power lines with heavy investments that relies on growth in revenues, we now should optimize the existing one with every flexibilities at our disposal. We also notice a revival of DC current technology, hybridizing the current AC grid with new dynamics. In addition, this new complexity also comes from other external factors such as the changing energy mix with a massive integration of renewables, as well as an ever more fragmented set of actors at a more granular level like prosumers, or at the supranational level with an interconnected European grid for instance.

This new complexity will bring new dynamics such as dynamically varying flow amplitudes and directions. This is in contrast of what was the case in the past with centralized production from large power plants, "pushing" the flows to the loads in a very hierarchical and descendant way. New distributed controls are getting implemented, taking advantage of new communication and software technologies. This pushes us towards an always more entangled cyber-physical system whose topology is no more the actual physical grid topology, which was convenient to study the grid. Its topology will be one also induced by long distance communications and controls. Therefore, rethinking the way we operate the grid has become a necessity. 

To handle the current complexity, our control room operators have built over time, and over many studies with the simulators at their disposal, their own mental representations of the grid. They actually segment the grid into static zones that are redefined every year to study the grid efficiently near real time.  They are indeed able to quickly identify remedial actions given security risks around. It helps them make the best trade-off between exploration and exploitation. However we anticipate that these yearly static views will be less and less relevant in the future to operate the grid, with fuzzier electrical "frontiers".  This can even occur along a course of a day within this dynamic context.
However a zonal segmentation should still be relevant to operate the grid by efficiently representing this complexity to act on it. Offering such context awareness will help our dispatchers in their decision making process. That is why an assisted segmentation built in a dynamic fashion to fit the specific context of a situation is needed.
Hence, how can we build such contextual segmentation for a given task? 

Previous works on segmentation have relied on the one hand on gathering proper dynamical phasor measurements on the grid to compute disturbance-based coherency in the time-domain and find similarities between electrical nodes
\cite{Kamwa2007Automatic,Wang1994Novel,Juarez2011Characterization}.  This implies a massive deployment of PMUs or very accurate large-scale dynamic simulations. On the other hand, other analytical approaches have investigated simplified modeling of the grid, relying on the linearized DC approximation, to partition it along buses for the purpose of studying cascading failures \cite{Blumsack2009Defining} (hierarchical clustering), \cite{Sanchez2014Hierarchical} (spectral clustering) \cite{Cotilla2013Multi} (hybrid K-means/evolutionary algorithm). This gave interesting results at a much lower cost. However, as our system becomes more cyber-physical with distributed regulations relying on advanced embedded software and fast communications, this has some limitations on the system complexity it can handle, such as non connected clusters in the actual grid topology.  In addition, given their  objective of identifying weak components overall for cascading failures, those methods were not particularly grid state specific. We would like to address those 2 points for our near-real time applications. 

In contrast of analytical methods that have been more extensively explored in the field of power systems, our approach relies on machine learning, following our previous work  \cite{IntroducingML} and responding to the call for new grid proxies in reliability management \cite{Proxies}.  We propose in this paper a new method that relies on a guided use of existing power grid simulators to teach the machine  an expected system response in a context of our task. We will talk of "guided machine learning", a form of unsupervised learning with carefully generated inputs to represent, guided by human expertise. For a more extended form of it, you can refer to \cite{GuidedML}.  Simulating systematic chosen interventions on a grid state, we build an Influence Graph (IG) to define a similarity between our components given our operational task. Interestingly, the IG connectivity goes beyond the actual grid topology, which can further lead to non topologically connected components within a cluster, an idea expressed by \cite{hines2015InfluenceG} and reminiscent of \cite{roy2001InfluenceModel} when studying cascading failures. This kind of phenomena will certainly become more prominent in a cyber-physical system and should be captured. Our machine can then learn a useful interpretable representation, a proxy, from that IG complex representation by running a suitable clustering algorithm. The \textit{Infomap} algorithm from the field of community detection was our top candidate given some intrinsic properties and has shown to work well on our IG.

The paper is organized as followed. Section \ref{sec:method} is dedicated to the method, where we describe the IG, justify its relevance compare to more classical distance matrices and talk about the suitability of the "\textit{Infomap}" clustering algorithm. In section \ref{sec:results} we present the results on commonly used system, namely the IEEE 14, for illustration and interpretation of our method. To compare our method to others, we use the well studied 96-RTS system for grid segmentation which can serve as a benchmark. We eventually give some insights on the usefulness of our method on large-scale power grids such has the French power grid.
Finally, section \ref{sec:conclusions} provides conclusions and future directions for this work.


\section{METHOD}
\label{sec:method}

A proxy, a simplified model of a complex system, can only be relevant for a certain range of tasks as we are "neglecting" details that matters for other phenomena. It is useful as it reduces the dimension and exploration of a problem related to our task while preserving the relevant information. Such a representation can be judged along 3 axis: interpretability (helping someone apprehend a situation), synthetic (limiting someone's exploration of the problem) and efficiency (containing solutions to the actual problem). 

Clustering methods already applies to a wide diversity of problems. Our main issue here is to provide representative data of our task for a given grid state to a clustering algorithm. Measurements are not enough as our grid state is evolving. The dynamics around this state are the results of multiple entangled phenomena whose contributions are hard to assess. We cannot invasively influence those dynamics on the real system for the sake of our method. Rather we need to rely on the proper use of simulators. Building on top of existing simulators has the advantage to rely on the complexity of their system modeling. This avoid us the need of redefining a specific analytical model that captures the grid complexity for our task. Furthermore, rather than analytically and explicitly modeling our task, we make use of the simulator as an oracle to show a system response under some representative experiments that we call interventions. We then let the machine learns from it a proper synthetic representation for this implicit task. The combination of the simulator (Sim) and our set of interventions peculiar to our task can be seen as a teacher for our machine: we will call it a guided simulator (GSim).

\subsection{An influence Matrix: a grid state under simulated small perturbations}
In this section, we suppose that we have at our disposal a simulator $Sim$, that given a grid state $(Inj,Topo)$, representing respectively the injections vector and the topology representation. The resulting call of this simulator is denoted by $x$:
\begin{equation}
x=Sim(Inj,Topo)
\end{equation}



Given our state $x = (\bm{z}, \omega)$, $\bm{z} = (z_1, \dots, z_j, z_{n_z})$ being our variables of interest relevant to our task, $\omega$ being the other variables we discard and with $(Inj,Topo)$  $\subset{x}$. In our case we have $\bm{z} = $ "all the active power flow on each line of the power grid", while neglecting voltage amplitude for example. These variables $\bm{z}$ are our variable of interest, and we want our clustering to best "represent" the "$\bm{z}$"'s complex interactions. To reveal this complexity, we can use a set of small perturbations $\mathcal{P} = \{ p_i \}_{1 \leq i \leq n_p}$ 
on either $Inj$ or $Topo$, around our grid state: $\forall p_i \in \mathcal{P} , p_i = \delta_{Inj}(x) \text{ or } \delta_{Topo}(x)$). For example $p_1$ can be "redispatching the production of the first generator of 1MW", and $p_2$ "disconnecting the $3^{\text{rd}}$ line of the power grid". We then run our simulator to assess the global effect $\Delta x_i$ of the perturbation $i$ on all the variables $x$:
\begin{equation}
\Delta x_i=|Sim( (Inj,Topo) \odot p_i)-Sim(Inj,Topo)|
\end{equation}
$\Delta x_i$ is the absolute difference between the state before the perturbation, and after it, which can see as the influence of $p_i$ on it. Recall that we are only interested in  the subset $\Delta \bm{z}_i$.


By stacking all the $\Delta \bm{z}_i$, we can define an Influence Matrix "IM" dedicated to our task (in our case studying the active flow $\bm{z}$) : 
\begin{equation}
IM_{i,j}=(\Delta \bm{z}_i)[j]_{{1 \leq i \leq n_p},{1 \leq j \leq n_z} }
\end{equation}
where $(\Delta \bm{z}_i)[j]$ denotes the $j^{\text{th}}$ value of the perturbated vector $\Delta \bm{z}_i$. 

To compute IM, we could imagine simulating any kind of perturbations, naively simulating for instance all possible injection redispatching actions and all topological changes, observing their effect on our $\bm{z}$ variables. We could then apply a clustering method on this entire cloud of observations.

However, computing all these perturbations is computationally too intensive  in practice and not always meaningful. We should then choose targeted and relatively non invasive small perturbations that leads to an overall system response, representative of our task. That's where the "guided" simulator comes into play. From(2), with carefully chosen set of perturbations $\mathcal{I}$, denoted as of now "intervention", we here define a guided simulator, given $(Inj,Topo) \subset{x}$:
\begin{equation}
\begin{split}
GSim(x,\mathcal{I})=|Sim((Inj, Topo)\odot \mathcal{I})-Sim((Inj, Topo))| 
\end{split}
\end{equation}



\subsection{An Influence Graph: guided simulations with systematic interventions }

A class of interventions we will use that is likely to match the above mentioned prerequisite are direct, independent and anaesthetizing interventions on each $z_i$. By a counterfactual reasoning, we can then observe how the system would have behaved without $z_i$ playing its role. This can thereby give us relevant information on the role of $z_i$ within the system, as well as highlights the close interactions with other $z_k$ that plays a similar role. Ideally we would like to decipher the causal influence of some injections and topologies on our $z_i$, which could help us gather explicitly $z_i$'s with common influent factors. This relates to the seminal work of \cite{Pearl}  on graphical models and causality. He indeed theorized the need for interventions in addition to observations to properly identify causal links and build proper graphical models. 

For this class of interventions $I_z$, we now have card($I_z$)=nz. The related IM is now no more than a square adjacency matrix of a directed graph. We define it as an Influence Graph IG. The nodes represents our variables z, the edges the interaction between $z_i$'s and the origin of an edge is given by the source $z_k$ on which the intervention $I_k$ occurred. To compute our contextual $IG_z$ for our task, we need to run our guided simulator, keeping a subset of the results related to z:

\begin{equation}
IG_z(x)=GSim_{x \in z}(x,I_z)
\end{equation}

Let's now apply this framework to our specific task of interest that is of importance for our operators today: managing the active power flows.

\subsection{The Influence Graph for managing active power flow}
For the task of managing active power flows, our $z$=$L^{pf}$ are the power flows on power lines of number $nL$. The intervention we choose {$I_{z_i}=Off(L^{pf}_i)$} is the disconnection of a given line to prevent the power flowing through it and hence setting it to 0. We hence observe how the system respond when redispatching the existing power flow through other pathways. This is a targeted and relatively non invasive intervention in regards of the active system variables, injections and topology. Indeed within a meshed transportation grid, it prevents power generation redispatching, maintaining the initial injection plan, and only modifying the topology locally and smoothly.

As a simulator, we use a static AC power flow $Sim=ACpf$ and hence compute our influence graph over all independent line disconnections $Off(z_i)$. Nodes in this graph correspond to our $z$, the power lines. An edge $e_{ij}$ has a weight:
\begin{equation}
w_{ij}=GACpf_{x \in L^{pf}}(x,Off(L^{pf}_i))_j, e_{ij}\in IG_{L^{pf}} . 
\end{equation}
A sensible threshold is set to remove numerical noise, the equivalent of 1MW on the French power grid which is often considered as a numerical tolerance.

\begin{figure}[!htbp]
\begin{center}
\includegraphics[width=3 cm, height=3.5cm]
{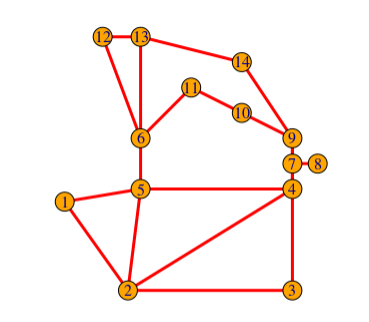}
\includegraphics[width=3 cm, height=3.5cm]
{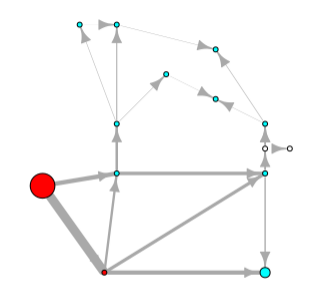}
\includegraphics[width=2.8cm, height=3.5cm]
{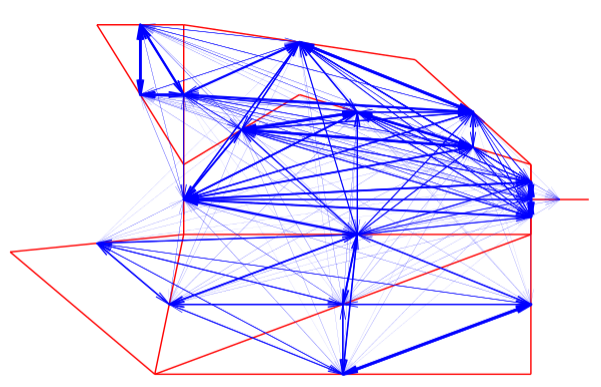}
\end{center}
\caption{3 representations of the grid. Topological on the right is the simplest. The power-flow one in the center adds the localization of injections, production in red and load in cyan, as well as the direction of flow. The Blue Influence Graph for power flows on the right eventually highlights the wide complexity of interactions between these flows through lines. An arrow goes from a line on which we intervene to a one it influenced.}
\label{fig:InfluenceGraph}
\end{figure}

Let's now visualize on Figure \ref{fig:InfluenceGraph} an $IG_{L^{pf}}$ over a relatively small system such as the IEEE 14 and compare it to other well-known representation of a grid, namely the grid topology and the grid power flows.

For the topological representation, we might want to segment the grid into 3 zones along the vertical axis when looking for cliques, that is coherently meshed zones. It misses  however the localization of injections. Considering them might actually lead to a different number of relevant zones. If productions and loads are distributed over all the nodes there might not be distinguishable sub-zones. If productions are concentrated in one part and loads in another as it is here, 2 zones could make more sense. 

Focusing now on the power-flow representation, it is more informative on that point. However it can be misleading as we will naturally follow the path of a given flow along its directions like a water flow, shadowing the superposition of interactions as stated by the superposition theorem. A power-flow is indeed a residual flow of multiple flows in both directions.  For flows not belonging to the same "water flow path", we will hence make independent zones whereas they could be strongly interacting. This difficulty doesn't arise clearly here as the production is quite concentrated and localized, hence pushing the flows along one path. We might want here to create 2 zones: a localized production zone with a clique of 3 in the bottom left, a diffuse consumption zone for the remaining grid. 

Now studying our $IG_{L^{pf}}$, it represents an additional level of complexity highlighting the superposition of interactions, some over long-distance. It shows the centrality of line 4-5 for our grid state as it is a bottleneck for our influence flows. We could expect from that observation to see our grid segmented into 2 electrically coherent zones with line 4-5 being the border line. 

If we can make that analysis on such a small grid, it might become quite impossible for a larger grid to interpret a wider IG. To synthesize this information and gain interpretability, we apply on it an unsupervised machine learning to help us better represent it. We hence choose and run an appropriate hierarchical community detection algorithm Infomap on our Influence Graph, resulting in various levels of representation for our task.

\subsection{Infomap: an information theory based community detection algorithm}
There are several algorithms for graph segmentation, known in literature as community detection algorithms \cite{Newman2002Finding,Guimera2004Modularity,Blondel2008Fast}. One can refer to the following article for a review on community detection algorithms \cite{Fortunato2010Community}.
The algorithm developed by Rosvall \textit{et al.} \cite{Rosvall2009map} known as "\textit{Infomap}", has the advantage of being particularly suitable for oriented, weighted graphs, and able to identify flow patterns inside the graph. It is recursively hierarchical \cite{Rosvall2011Multilevel} and can automatically find the proper number of hierarchical levels and clusters. In addition, "\textit{Infomap}" can handle overlapping  \cite{Viamontes2011Compression} which could be of interest for future works. Indeed electrical frontiers are fuzzy and it could make sense that lines interconnect some clusters. We will here briefly describe the main ideas of this method, the reader can refer to the original article for a complete description.

The idea is to use the duality between the problem of how we should best partition the network with respect to flow and the problem of minimizing the description length of places visited along a path given the influence graph.
The goal is hence to compress the information by making the best encoding to name our variables: a codeword.
This is all goes back to Shannon's source coding theorem  from information theory which establishes that for $n$ codewords describing $n$ states of a random variable $X$ that occur with frequencies $p_i$, the average length of a codeword can be no less than:  $H(X) = - \sum ^{n} _{1} p_i log(p_i)$. The more you use a codeword, the smallest encoding you should give it. To better minimize the average encoding length of codewords, one can further take advantage of the graph regional cycling structure that highlights modules $M$, and define a "module codebook" for each area that contains all the nodes codewords of this area. Thereby it is possible to reuse the same codeword for different nodes since we can specify which module codebook to use. We then need an "index codebook" containing a codeword for each "module codebook". Going from one node to another in the same region, one only needs to refer to a short codeword to identify it, knowing the region codebook. An easy analogy is the case of maps with streets, cities and countries. For instance, in different cities you will find the same street names, and you can do so because you can name the city as well, to better identify this street in a country. But being in a given city, you don't even need to name the city again to refer to a street in it: you compressed the information while still being able to communicate it.

To eventually minimize the description length of our variables, we can recursively apply Shanon's theorem to codewords and codebooks which leads to the map equation:

\begin{equation}
L(M) = q_{\curvearrowright} H(Q) + \sum _{i=1} ^m p_{\circlearrowright} ^{i} H(P^i) 
\end{equation}

with $H(Q)$ the weighted average length of codewords in the codebook index and $H(P_i)$, the weighted average length of codewords in the module codebook $i$. The codeword index is used at frequency $q_{\curvearrowright}$ the probability to change module at every step of the random walk. The $i$ module codebook is used at frequency $p_{\circlearrowright} ^{i}$, which is the number of moves  continually spent in module i plus the probability to leave.
In practice to compute the frequencies, they use a random walker over the graph.

\section{Results}
\label{sec:results}
We applied our method on system of different sizes to visualize the  method genericity. First we illustrate it on the reduced IEEE-14. We then benchmark our method on the 96-RTS system which has been studied in the past as an interesting baseline for segmentation purposes. We further share results on the 118-bus system which is a realistic and readable middle-sized one. We finally show a segmentation on the large-scale French power grid to demonstrate how it scales while being able to give initial interpretations.

\subsection{IEEE-14}
The IEEE-14 is an appropriate system to illustrate our method. From figure 2, we see that the system has a production zone and a distribution one. Our flows are highly influenced by one or the other. We expect our method to segment this system into 2, a cluster for productions and another for loads. This is what we observe on our results, with the line "4\_5" being a border. Our representation is hence interpretable since it helps understanding the structure of the power system. 

\begin{figure}[!htbp]
\begin{center}
\includegraphics[width=3 cm, height=3cm]
{InfluenceGraphIEEE14.png}
\includegraphics[width=3 cm, height=3cm]
{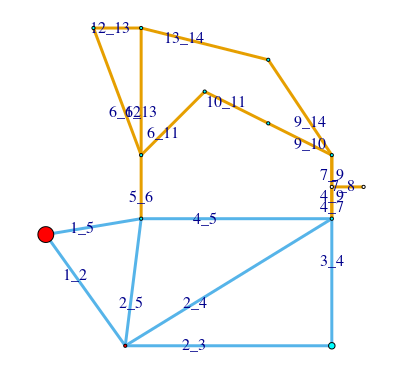}
\caption{On the left, our IG for the IEEE 14 system. On the right, our segmentation into 2 zones using InfoMap on our IG}
\end{center}
\label{fig:ieee14}
\end{figure}




\subsection{IEEE-RTS-96}
\begin{figure}[!htbp]
\begin{center}
\includegraphics[width=7cm, height=3.5cm,trim={1cm 3cm 2cm 2.5cm},clip]{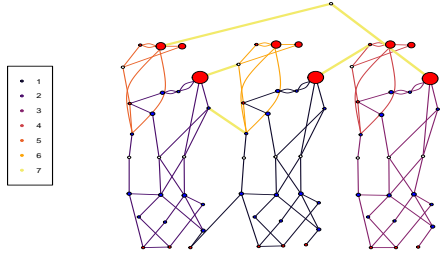}
\end{center}
\caption{IEEE-RTS-96 segmentation. Production in red and load in blue, sized by volume. The 7th non-connected cluster is in yellow. }
\label{fig:ieee96}
\end{figure}

To benchmark our method, we used the reliability test system 1996 from which we obtained the clustering showed in figure \ref{fig:ieee96}.
It highlights one level and 7 clusters, 6 agreeing with the power grid connectivity and 1 not. We argue that this surprising non-connected cluster comes from the system and is not an artifact of our method. This will be discussed after. As for the 6 others, since they represent the same IEEE-24 case
,it is consistent to segment them in the same way.  

\begin{figure}[!htbp]
\begin{center}
\includegraphics[width=5cm, height=3cm]{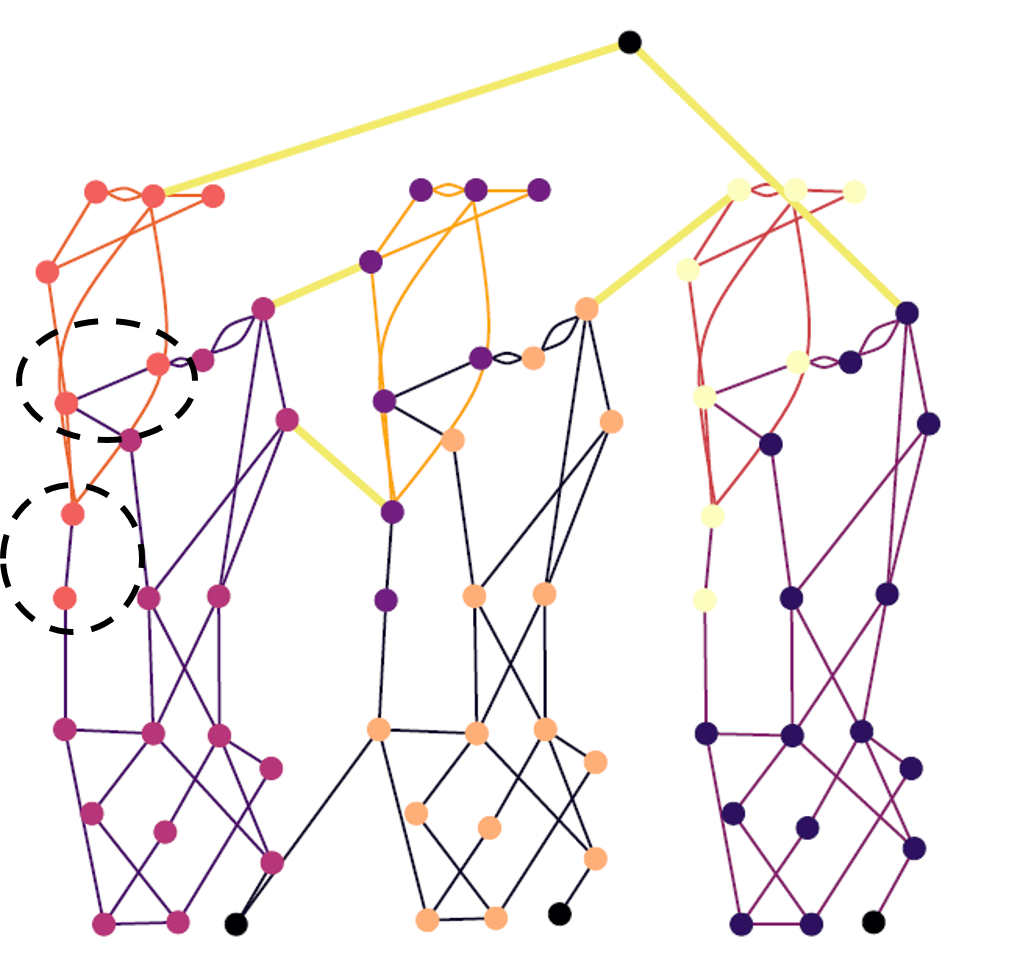}
\end{center}
\caption{Comparison of our IEEE-RTS-96 segmentation (line colored) with two other methods (node colored)\cite{Cotilla2013Multi}\cite{Kamwa2007Automatic}. Recurrent dissemblances highlighted on 1 subgrid, plus the non-connected cluster}
\label{fig:ieee96compa}
\end{figure}

We compare the results of our method to other works in figure \ref{fig:ieee96compa}, \cite{Cotilla2013Multi} who uses electrical distance and \cite{Kamwa2007Automatic} who uses time-domain measurements. Overall, the clusterings are very similar while being computed by 3 different methods which might indicate we are close to an interesting and useful clustering. 

There are however slight differences we can comment on. We can notice 2 differences as circled on the figure \ref{fig:ieee96compa}, besides that we are actually clustering lines and they are clustering nodes. 
Topologically speaking, we argue that our method properly circumscribes the upper cluster to a meshed clique whereas the other clustering has slight less obvious unmeshed extensions.

About the non-connected one, it gathers high voltage interconnecting power lines close to productions that interconnects the three same sub-grids, but leave aside the low voltage interconnection close to loads, which can be understood. We rediscovered that that these lines were artificial additions for the purpose of this synthetic system, and not the result of a coherent grid development. Cutting one of those power lines leads to significant changes in flows over the whole grid, 
as illustrated in the Influence Graph heatmap in figure \ref{fig:heatMap AS}. Hence these interconnecting lines play the same role in the power grid, 
even if they are non-connected, and it then makes sense to cluster them together.

\subsection{Beyond the grid topological graph}
\begin{figure}[!htbp]
\begin{center}
\includegraphics[width=4.5 cm, height=3cm]
{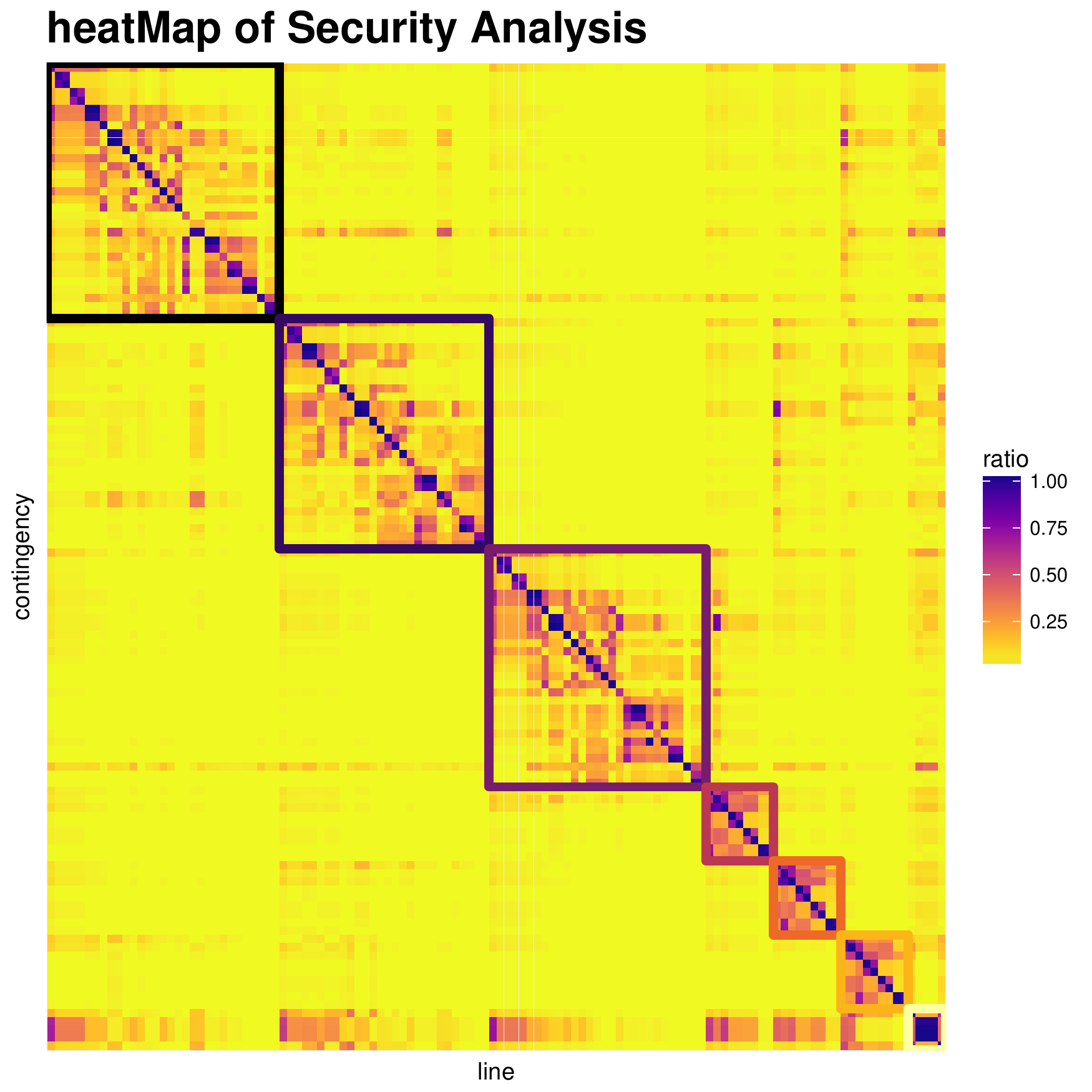}
\includegraphics[width=4cm, height=3cm]
{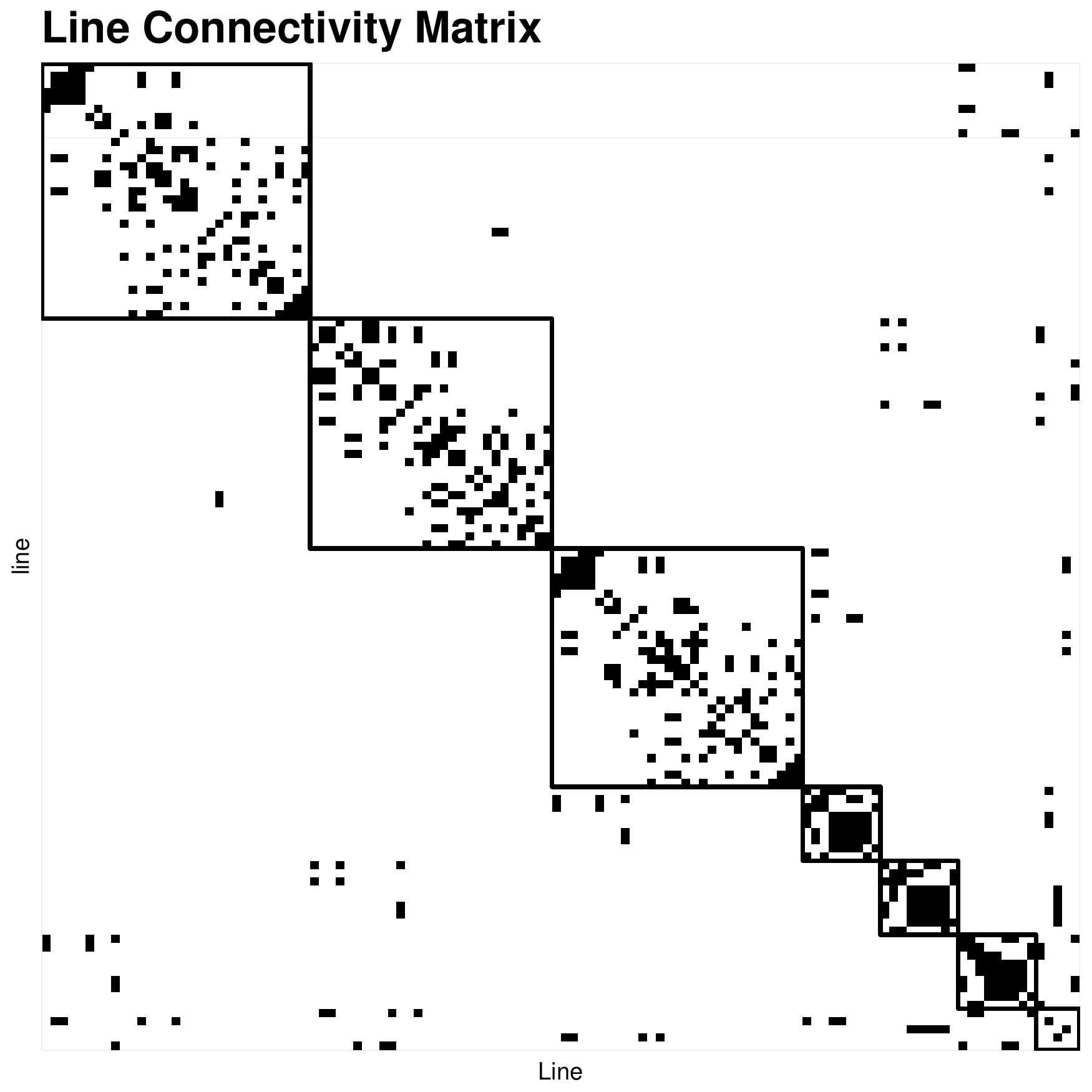}
\end{center}
\caption{HeatMaps to illustrate graph connectivity given different representations for RTS-96 system:  a) Influence Graph b) topological graph. The bottom rows of those graph matrices are filled quite differently and represent the non topologically connected cluster in the Influence Graph.}
\label{fig:heatMap AS}
\end{figure}

Very little works for grid segmentation have tried to use representations of the grid that go beyond its connectivity, a hard constraint which seems at first natural and intuitive. Nevertheless, this might overlook that a power system is a very entangled one with complex interactions, sometimes counter-intuitive as \cite{hines2015InfluenceG} explained. Here we use an influence graph that has a different connectivity than the grid as shown in figure \ref{fig:heatMap AS} for RTS-96 system \cite{Grigg1999IEEE}: it is actually more connected. But in the segmentation process, some links will appear stronger and relevant while other will be weak and ignored. As a consequence, the results will most generally lead to connected elements in clusters from the grid topological graph perspective. But some might not be connected, which could highlight complex interactions between flows and potential areas where the grid needs to be reinforced. This is one of the interesting interpretation we retrieve when applying our method.

To confirm this primary analysis on the graphs, we run Infomap on classical topological representation of the grid which both resulted in a 3-zone segmentation, missing our interconnection line cluster and not representative of the localization of injections.    

\begin{figure}[!htbp]
\begin{center}
\includegraphics[width=4cm, height=3cm,trim={5cm 3cm 5cm 3cm},clip]{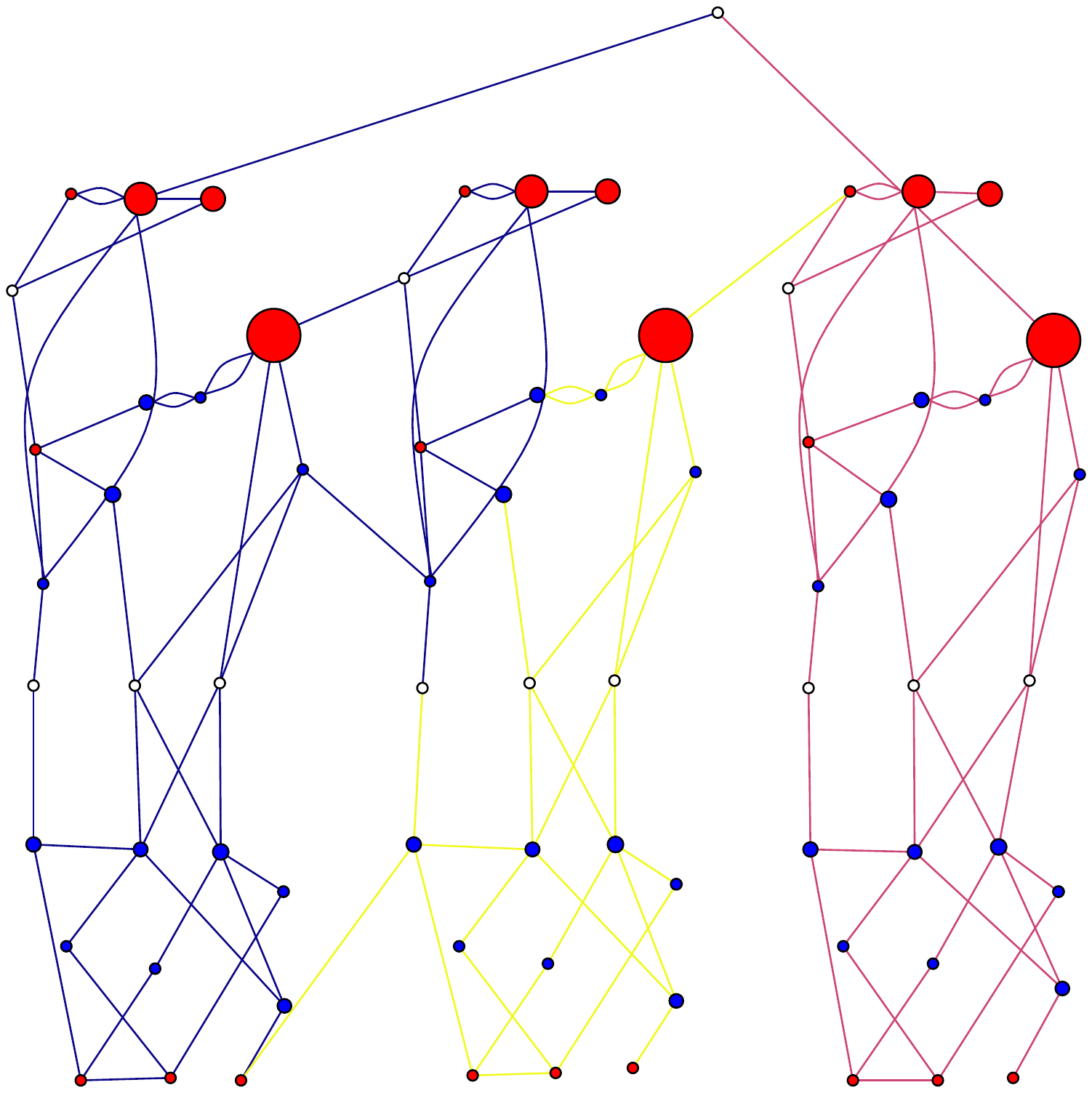}
\includegraphics[width=4cm, height=3cm,trim={5cm 3cm 5cm 3cm},clip]{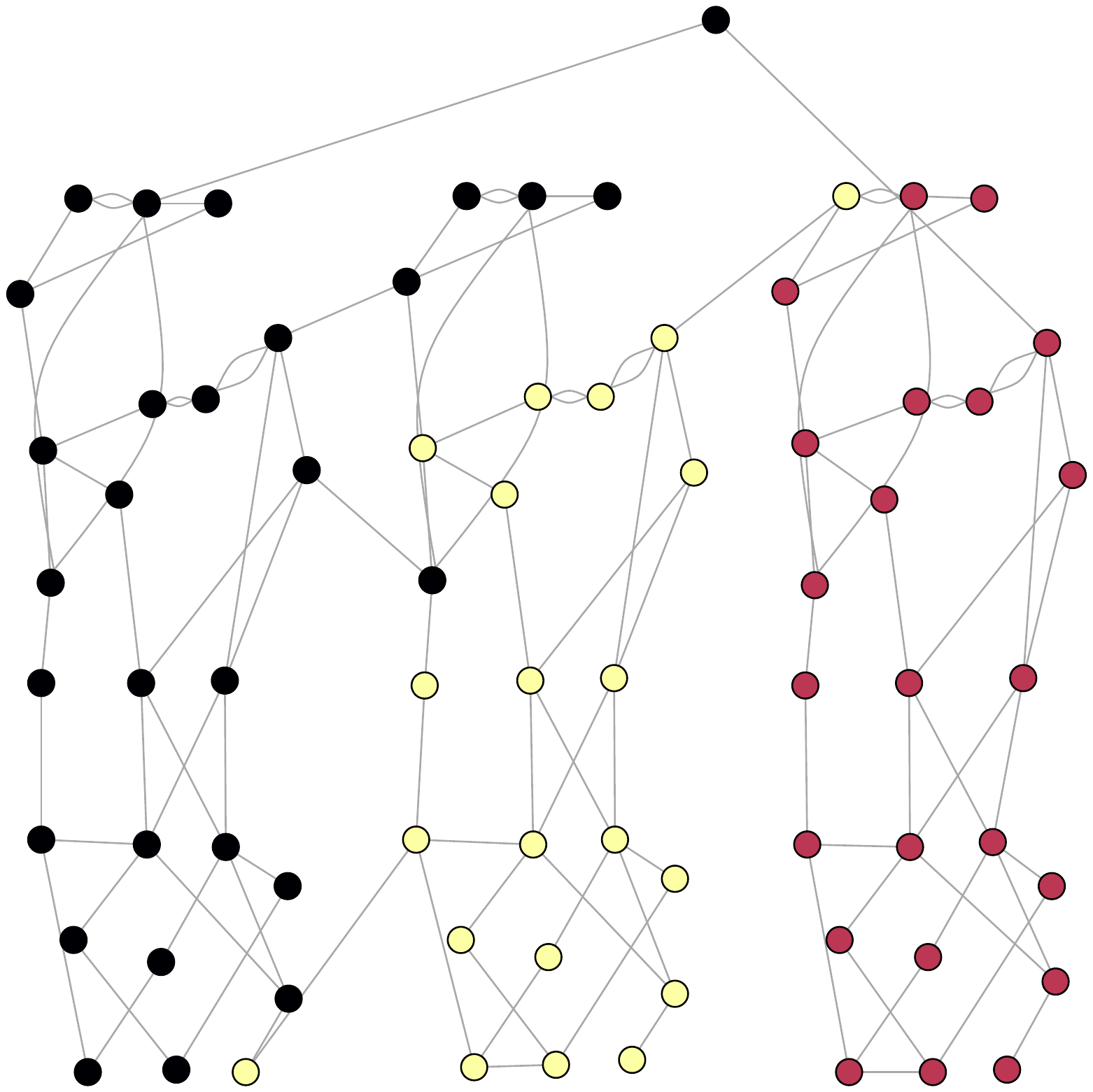}
\end{center}
\caption{Naive IEEE-RTS-96 segmentation with: a) connectivity matrix b) conductance matrix.} 
\label{fig:ieee96-naiveApproach}
\end{figure}

This is an example of one possible other application of our method: identifying weakly meshed interconnecting areas that are strongly interacting over long distance. We can hence capture clusters that are non-connected in the actual grid topology.

\subsection{Middle-sized grid: IEEE-118}



\begin{figure}[!htbp]
\begin{center}
\includegraphics[width=6cm, height=3.5cm,trim={1.5cm 3cm 5cm 2cm},clip]{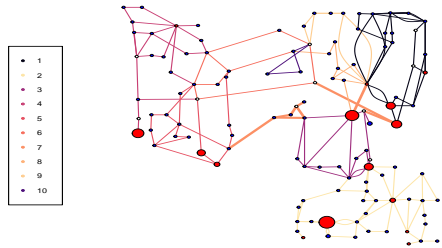}
\end{center}
\caption{IEEE-118 segmentation. The orange non-connected cluster 7 with East-West interconnections is highlighted with thick edges.}
\label{fig:ieee118}
\end{figure}

The IEEE-118 bus test case is a reduced model of the Midwestern US power grid in 1962 \cite{PSTCA2001Power}.
In figure \ref{fig:ieee118}, one can see the IEEE-118 case segmentation. We can distinguish at the top level 9 clusters, 8 agreeing with the grid connectivity and 1 which does not. This non-connected cluster 7 in the grid topology plays the same role as its counterpart in the IEEE-96-RTS case: important weakly-meshed interconnections between two well-meshed East and West grids, with some unbalance here, the East grid having more production and the West too much consumption. The remaining connected clusters seem reasonable with a proper clique segmentation and localized injections. We could actually expect some overlapping for some lines in-between clusters so that they each have proper cliques. This is something we actually observe when running \textit{Infomap} with that option and will be further studied in future works.

\vspace{-5pt}
\subsection{Large grids: the French power grid}

\begin{figure}[!htbp]
\begin{center}
\includegraphics[width=4 cm, height=3cm]
{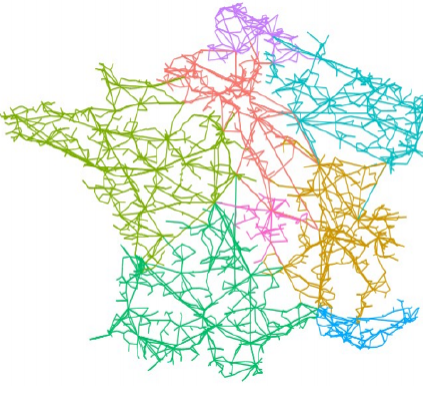}
\includegraphics[width=4cm, height=3cm]
{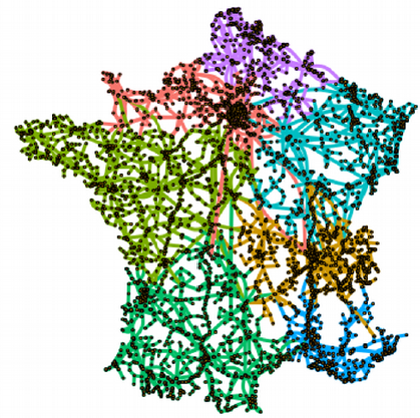}
\end{center}
\caption{Comparison of a) our French power grid segmentation with b) historical RTE regional segmentation. }
\label{fig:reseauRTE}
\end{figure}

Finally, we discuss our segmentation over the French power grid composed of $6~000$ nodes, $10~000$ lines and $10~000$ injections on a snapshot 19th January 2012 at 7pm. Running our method we actually find 8 clusters for the level 1 of the hierarchical clustering. 8 is here an result of Infomap, not a prior input parameter. It is remarkably close to 7, the number of RTE historical regional segmentation. We decided to compare those 2 segmentations and figured out they were actually quite close as you can see visually on figure \ref{fig:reseauRTE}. The historical RTE partition is not a pure electrical one, human resources, workload, maintenance teams and their localization were taken into account by that time. Nonetheless, this preliminary result is very encouraging as we can give it such interpretation and it has been positively commented by our dispatchers. Further validation on the quality of lower level clusters such as level 2, that have synthetic sizes close to the areas usually drawn by our dispatchers, could lead to redesigning some of our study tools, offering them context awareness in a more dynamical cyber-physical system.    

\vspace{0pt}


\section{CONCLUSIONS}
\label{sec:conclusions}
In this paper we derived a new method to efficiently segment power grids. The method relies on a guided machine learning approach build on top of existing physical simulators. We applied it to the task of studying power flows on a grid state and the resulting synthetic segmentations led to a successful benchmarking and meaningful interpretations. In particular it highlights non-connected clusters in the grid topology illustrating the grid complexity. It also finds itself an appropriate number of clusters. We believe that our approach could generalize to more cyber-physical systems and could be extended to create other meaningful representations for other tasks of interest. Future works will define more quantitative measures beside our interesting analytical results, to further validate our unsupervised method. New analysis will also be conducted  on lower level clusters, on overlapping, on clustering evolution over time, to eventually assess its efficiency and the importance of grid state context. Our method could then become a building block for new contextual visualization of the power system or for targeted control applications with reduced computation. 
\addtolength{\textheight}{-12cm}   
\bibliographystyle{unsrt}
\bibliography{bibliography.bib}

\end{document}